\documentclass[aps,pra,twocolumn,footinbib,superscriptaddress]{revtex4-1}
\usepackage{graphicx}
\usepackage{wrapfig}
\usepackage{indentfirst}
\usepackage{braket}
\usepackage{float}
\usepackage{amsmath}

\newcommand{\bk}{{\bf k}}
\newcommand{\ba}{{\boldsymbol a}}

\newcommand{\bLam}{\boldsymbol \Lambda}
\usepackage{bm}

\begin{document}

\title{The Quantum Illumination Story}

\author{Jeffrey H. Shapiro}
\email{jhs@mit.edu}
\address{Research Laboratory of Electronics\\
Massachusetts Institute of Technology\\
Cambridge, Massachusetts 02139, USA}

\begin{abstract}
Superposition and entanglement, the quintessential characteristics of quantum physics, have been shown to provide communication, computation, and sensing capabilities that go beyond what classical physics will permit.  It is natural, therefore, to explore their application to radar, despite the fact that decoherence---caused by the loss and noise encountered in radar sensing---destroys these fragile quantum properties.  This paper tells the story of ``quantum illumination'', an entanglement-based approach to quantum radar, from its inception to its current understanding.  Remarkably, despite loss and noise that destroy its initial entanglement, quantum illumination does offer a target-detection performance improvement over a classical radar of the same transmitted energy.  A realistic assessment of that improvement's utility, however, shows that its value is severely limited.  Nevertheless, the fact that entanglement can be of value on an entanglement-breaking channel---the meta-lesson of the quantum illumination story---should spur continued research on quantum radar. 
\end{abstract}
 
\maketitle
 
\section*{Introduction} 
Superposition and entanglement---Schr\"{o}dinger's cat being simultaneously alive and dead~\cite{Schroedinger1935}, and the ``spooky action at a distance'' that Einstein found disturbing~\cite{EPR1935}---are quantum-mechanical phenomena that are moving from fundamental studies into scientific and engineering applications.  Quantum computers, if realized at sufficiently large scale, will vastly outstrip the capability of classical machines for a variety of problems in simulation~\cite{Georgescu2014}, optimization~\cite{Boixo2014}, and machine learning~\cite{Biamonte2017}.  Those large-scale quantum computers---running Shor's quantum factoring algorithm~\cite{Shor1994}---will also break the public-key infrastructure on which Internet commerce currently relies.  Quantum communication, in the form of quantum key distribution~\cite{BB84}, however, may thwart that  quantum threat.  In other work, quantum-enhanced sensing is moving out of the laboratory and into real use, the most notable example being the incorporation of squeezed-state light into the laser interferometric gravitational-wave observatory (LIGO)~\cite{LIGO2013}.   A natural question to ask, therefore, is whether quantum techniques can bring performance gains to radar sensing.  This paper will follow one avenue of quantum radar research from its inception to now:  quantum illumination.   

\section*{Lloyd's Quantum Illumination}

\begin{figure*}[hbt]
\centering
\includegraphics[width=6in]{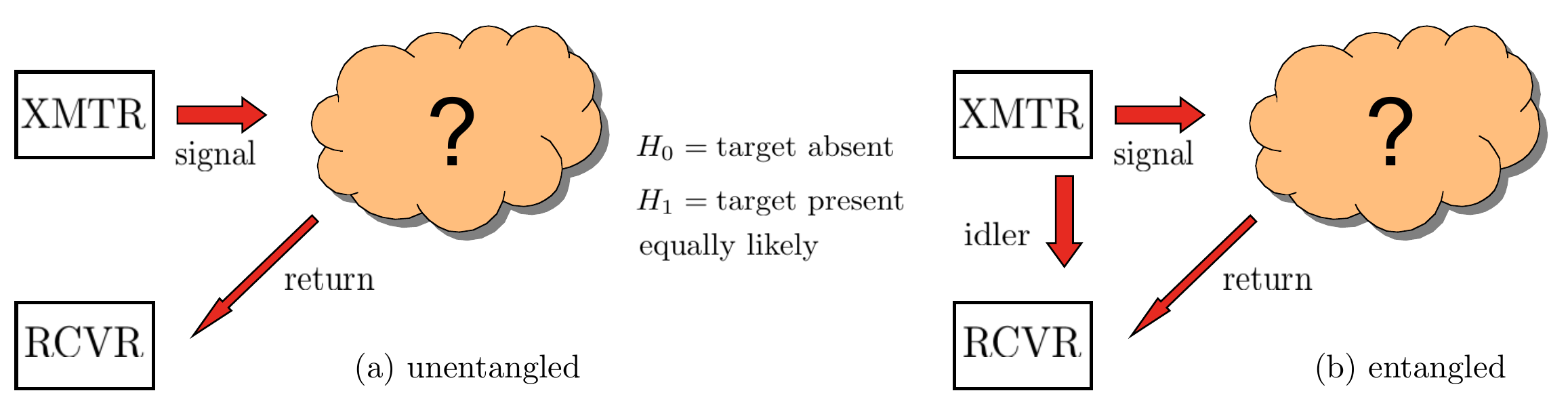}
\caption{Scenarios presumed in Lloyd's treatment of quantum illumination (QI).  (a) Unentangled operation, in which the transmitter illuminates the region of interest with a sequence of $N$ single-photon signal pulses.  Target absence or presence is decided from observation of the returned light. (b) Entangled operation, in which the transmitter illuminates the region of interest with a sequence of $N$ single-photon signal pulses, each of which is entangled with a companion single-photon idler pulse.  Target absence or presence is decided from observation of the retained idler and the returned light.  
\label{QI_fig1}}
\end{figure*}

Lloyd~\cite{Lloyd} coined the term ``quantum illumination'' for his entanglement-based approach to improving an optical radar's capability to detect a weakly-reflecting target embedded in background noise that can be much stronger than the target return.  His work, which built on Sacchi's earlier studies of quantum operation discrimination~\cite{Sacchi1}, \cite{Sacchi2}, compared the target-detection performance for the two scenarios shown in Fig.~\ref{QI_fig1}.  In both scenarios an optical transmitter illuminates a region of space in which a weakly-reflecting target is equally likely to be absent or present within always-present background light.  In Fig.~\ref{QI_fig1}(a), the transmitter's signal beam is a sequence of $N$ high time-bandwidth product ($M = TW \gg 1$), single-photon pulses.  The receiver, for this single-photon (SP) scenario, makes a minimum error-probability decision between hypotheses $H_0$ (target absent) and $H_1$ (target present) from observation of the light returned from the interrogated region.  In Fig.~\ref{QI_fig1}(b), the transmitter illuminates the region of interest with a sequence of $N$ high time-bandwidth product ($M = TW \gg 1$), single-photon signal pulses, each of which is entangled with a companion single-photon idler pulse; see Appendix~\ref{AppA} for the details.  The receiver, for this quantum illumination (QI) scenario, makes its minimum error-probability decision between $H_0$ and $H_1$ from observation of the retained idler light and the light returned from the interrogated region.

The crucial assumptions in Lloyd's analysis of the unentangled and entangled scenarios are as follows.
\begin{itemize}
\item When the target is present, the roundtrip transmitter-to-target-to-receiver transmissivity for the signal beam is $0 <\kappa \ll 1$.
\item The background light's average photon number per temporal mode, $N_B$, satisfies the low-brightness condition $N_B \ll 1$.
\item For each transmitted signal pulse, at most one photon is returned to the receiver, regardless of whether the target is absent or present, implying that $MN_B \ll 1$.  
\end{itemize}
Under these assumptions, Lloyd identified two operating regimes, the ``good'' and the ``bad'', for his SP and QI scenarios, and compared their error probabilities' quantum Chernoff bounds~\cite{footnote1} in these regimes.   In their good regimes, SP and QI's error probabilities have the same bound,
\begin{equation}
\Pr(e)_{\rm SP} \le e^{-N\kappa}/2 \quad\mbox{and}\quad \Pr(e)_{\rm QI} \le e^{-N\kappa}/2,
\label{LloydGood}
\end{equation}
which, because $N\kappa$ equals the average number of signal photons returned when the target is present, whereas 0 is the average number of signal photons returned when the object is absent, equals the signal shot-noise limit for laser communication with on-off-keying (OOK) modulation.  Despite (\ref{LloydGood})'s applying to SP and QI's good regimes, QI still enjoys a substantial good-regime performance advantage over SP, because SP's good regime is limited to $\kappa \gg N_B$, whereas QI's extends to the much larger parameter region in which $\kappa \gg N_B/M$. 

The comparison between SP and QI operation is radically different in their bad regimes.  Here Lloyd found that
\begin{equation}
\Pr(e)_{\rm SP} \le e^{-N\kappa^2/8N_B}/2, \mbox{ for $\kappa \ll N_B$},
\label{LloydSPbad}
\end{equation}
and
\begin{equation}
\Pr(e)_{\rm QI} \le e^{-N\kappa^2M/8N_B}/2, \mbox{ for $\kappa \ll N_B/M$}.
\label{LloydQIbad}
\end{equation}
These bounds mimic the background-limited error probability of OOK laser communication, with $N_B$ being the background's brightness for SP operation, and $N_B/M$ being that brightness for QI operation.  
The bad-regime results reveal a double advantage for QI:  first, its bad regime applies in a smaller $\kappa$ range than that of SP; and second, when both systems are in their bad regimes, QI enjoys a factor-of-$M$ greater error-probability exponent (effective signal-to-noise ratio).  At optical frequencies, high time-bandwidth product is easy to obtain, e.g., a 1\,$\mu$s pulse at 300\,THz center frequency (1\,$\mu$m wavelength) with 1\,THz (1/3-percent fractional) bandwidth yields $M = 10^6$, in which case Lloyd's bad-regime QI has a 60\,dB higher effective signal-to-noise ratio than its SP competitor.  Moreover, and remarkably, this advantage is afforded despite the background noise's destroying the initial entanglement, i.e., the retained and returned light are not entangled. 

Despite the idealized nature of Lloyd's QI analysis---it presumes an on-demand source of high-$TW$ entangled signal-idler photon pairs, lossless idler storage, and perfect realization of an optimum quantum receiver---its enormous predicted performance enhancement in the bad regime motivated a great deal of follow-on research.  Some of that research, unfortunately, took much of the air out of QI's balloon, as we will describe below.  Before doing so, however, a brief preface about classical versus quantum radar is in order.    

We were careful, earlier in this section, \emph{not} to describe the comparison between SP and QI operation as one between a classical radar (SP) and a quantum radar (QI).  In quantum optics, see, e.g.,~\cite{Shapiro2009}, \cite{Shapiro2012}, it is conventional to reserve the appellation ``quantum'' for those systems whose performance analysis \emph{requires} the quantum theory of photodetection.  In particular, their performance cannot be correctly quantified from the semiclassical theory of photodetection, in which light is treated as a classical (possibly stochastic) electromagnetic wave and the discreteness (quantization) of the electron charge gives rise to shot noise.  Furthermore, measurements of light beams that are in coherent states~\cite{Glauber1963}, or classically-random mixtures thereof, in any of the three basic photodetection paradigms---direct detection, homodyne detection, or heterodyne detection---do \emph{not} require quantum photodetection theory to obtain correct measurement statistics~\cite{Shapiro2009}.  Hence such states are called classical states.  Systems that employ \emph{nonclassical} states---states other than coherent states or their random mixtures---require the use of quantum photodetection theory.   In this regard, we note that single-photon states and entangled states are not classical states, and hence Lloyd's performance comparison is between two quantum radars, one of which employs entanglement (QI) while the other (SP) does not. 

A coherent state---other than the zero-photon (vacuum) state---contains a random number of photons.  So, Shapiro and Lloyd~\cite{ShapiroLloyd} compared Lloyd's QI to a classical radar by replacing Lloyd's SP transmitter in Fig.~\ref{QI_fig1}(a) with a coherent-state transmitter (an ideal laser) that produces a sequence of $N$ pulses, each of which has unity \emph{average} photon number.  The quantum Chernoff bound for their coherent-state radar,
\begin{equation}
\Pr(e)_{\rm CS} \le e^{-N\kappa(\sqrt{1+N_B}-\sqrt{N_B})^2}/2,
\label{ShapiroLloydCS}
\end{equation}
which applies for all $0\le \kappa \le 1$ and for all $N_B \ge 0$, reduces to 
\begin{equation}
\Pr(e)_{\rm CS} \le e^{-N\kappa}/2,
\end{equation}
for the low-brightness ($N_B \ll 1$) background that Lloyd's QI analysis assumed.  Shapiro and Lloyd thus showed that a coherent-state radar matched the performance of Lloyd's QI radar in the latter's good regime, and was substantially better when QI operated in its bad regime.  Fortunately for quantum illumination, Shapiro and Lloyd's work was not the end of the story; Tan~\emph{et al}.~\cite{Tan} had already analyzed a Gaussian-state QI system that outperformed \emph{all} classical radars of the same transmitted energy.

\section*{Tan~\emph{et al}.'s Quantum Illumination}
\begin{figure*}[hbt]
\centering
\includegraphics[width=6in]{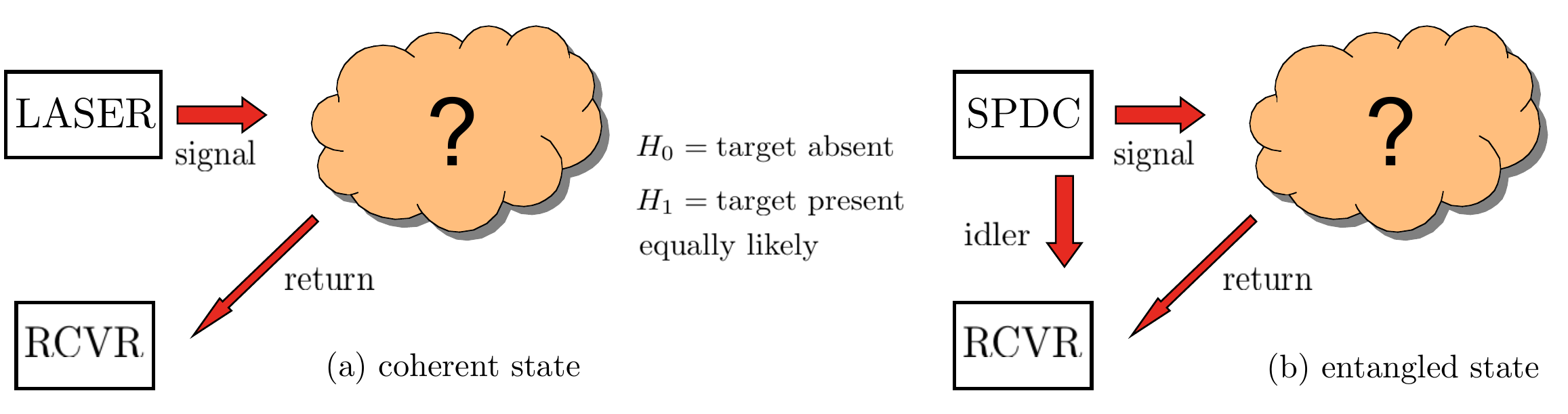}
\caption{Scenarios presumed in Tan~\emph{et al}.'s treatment of quantum illumination (QI).  (a) Coherent-state operation, in which a laser transmitter illuminates the region of interest with a coherent-state signal pulse of average photon number $MN_S$.  Target absence or presence is decided from observation of the returned light. (b) Entangled operation, in which the transmitter illuminates the region of interest with a duration-$T$ signal pulse carved from the low-brightness (average photon number per temporal mode $N_S \ll 1$) output of a continuous-wave spontaneous parametric downconversion (SPDC) source whose phase-matching bandwidth $W$ satisfies $M= TW \gg 1$.  Target absence or presence is decided from observation of the retained idler and the returned light.  
\label{QI_fig2}}
\end{figure*}

Tan~\emph{et al}.~compared the classical and quantum radar scenarios, shown in Fig.~\ref{QI_fig2}, for detecting a weakly-reflecting ($0<\kappa \ll 1$) target that is equally likely to be absent or present within always-present, high-brightness ($N_B \gg 1$) background light.    The classical radar, in Fig.~\ref{QI_fig2}(a), illuminates the region of interest with a coherent-state (laser) pulse of average photon number $MN_S$, where $N_S \ll 1$ and $M = TW \gg 1$.  Target absence or presence is then decided from observation of the returned light.  The QI radar, in Fig.~\ref{QI_fig2}(b), carves duration-$T$, entangled signal and idler pulses from the outputs of a continuous-wave spontaneous parametric downconversion (SPDC) source, whose phase-matching bandwidth is $W$, and whose signal and idler have average photon number per temporal mode $N_S \ll 1$; see Appendix~\ref{AppB} for the details.  Target absence or presence is then decided from observation of the retained idler and the returned light. 

Pirandola and Lloyd~\cite{Pirandola2008} developed a formula for computing the quantum Chernoff bound for the task of distinguishing between two arbitrary multi-mode Gaussian states.  Exploiting this tool, Tan~\emph{et al.} showed that the quantum Chernoff bounds for their coherent-state (CS) and QI radars behave rather differently than those for Lloyd's SP and QI radars, i.e., Tan~\emph{et al}.~found
\begin{equation}
\Pr(e)_{\rm CS} \le e^{-M\kappa N_S/4N_B}/2,
\label{TanCS}
\end{equation}
and 
\begin{equation}
\Pr(e)_{\rm QI} \le e^{-M\kappa N_S/N_B}/2,
\label{TanQI}
\end{equation}
where, in both cases, $0<\kappa \ll 1$, $N_S \ll 1$, and $N_B \gg 1$ are assumed.  By analogy with Lloyd's work, we might term this operating regime the ``bad'' regime for Tan~\emph{et al}.'s CS and QI systems.  Lloyd's bad-regime QI system offered an $M$-fold error-probability exponent improvement over its SP counterpart, where a time-bandwidth product $M \ge 10^6$ is easily attainable.  In contrast, Tan~\emph{et al}.'s QI system only affords a factor-of-four (6\,dB) improvement in error-probability exponent over its CS counterpart of the same transmitted energy, \emph{regardless} of an $M \ge 10^6$ time-bandwidth product.  Moreover, because Tan~\emph{et al}.~do not restrict their $M$-mode signal and idler state spaces to the span of their $M$-mode vacuum and single-photon states, their work does \emph{not} fall prey to issues identified by Shapiro and Lloyd.  

\begin{figure}
\centering
\includegraphics[width=2.5in]{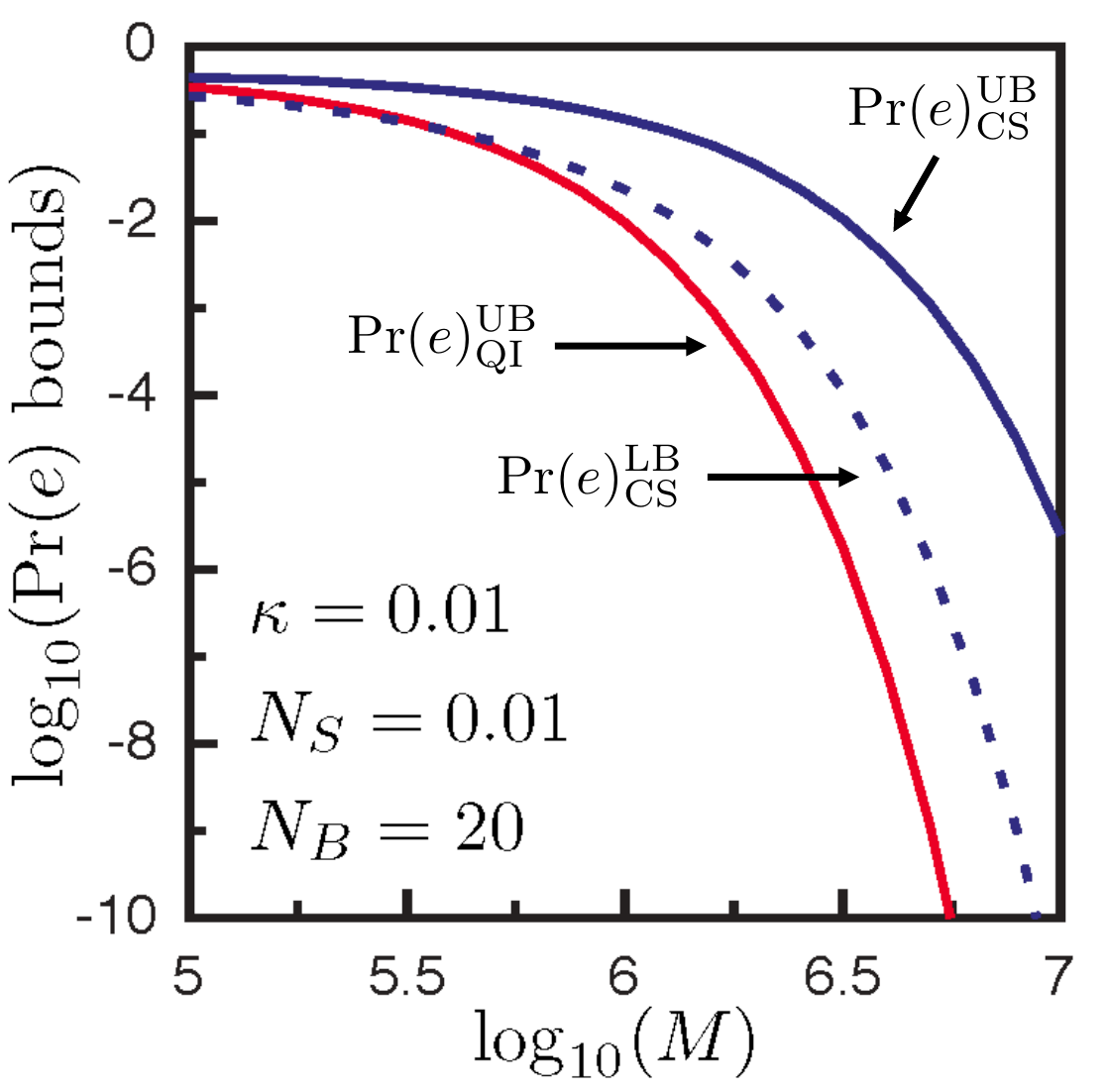}
\caption{Error-probability bounds for Tan~\emph{et al}.'s CS and QI radars~\cite{Tan}:  $\Pr(e)_{\rm CS}^{\rm UB}$ is the quantum Chernoff bound for the CS radar;  $\Pr(e)_{\rm QI}^{\rm UB}$ is the quantum Chernoff bound for the QI radar; and $\Pr(e)_{\rm CS}^{\rm LB}$ is the Bhattacharyya lower bound for the CS radar. 
\label{QI_fig3}}
\end{figure}

Figure~\ref{QI_fig3} illustrates the behaviors of Tan~\emph{et al}.'s quantum Chernoff bounds for $\kappa = 0.01$, $N_S = 0.01$, and $N_B = 20$.  Also included in this figure is the Bhattacharyya \emph{lower} bound on $\Pr(e)_{\rm CS}$ from Ref.~\cite{Tan}.  It is important to remember that quantum Chernoff bounds are known to be exponentially-tight upper bounds, while it is known that the Bhattacharyya lower bound is always loose.  Furthermore, Tan~\emph{et al}.~showed that the CS radar affords the lowest error probability of any classical-state radar of the same average transmitted energy.     Thus, because Fig.~\ref{QI_fig3} shows $\Pr(e)_{\rm QI}^{\rm UB}$ becoming  lower than $\Pr(e)_{\rm CS}^{\rm LB}$ for sufficiently high $M$ values, it provides definitive proof that a QI radar can, in principle, outperform \emph{all} classical radars of the same transmitted energy for the target-detection scenario addressed by Tan~\emph{et al}.  We will say more later about why we refer to Fig.~\ref{QI_fig3}'s QI advantage over CS operation as being ``in principle''. For now, it suffices to point out that when Ref.~\cite{Tan} appeared there was no known receiver that provided \emph{any} QI performance advantage over CS operation.   Why that was so and how it was overcome require some understanding of quantum photodetection theory and the quantized electromagnetic field's Gaussian states.  So those topics are next on our agenda.

The $m$th temporal modes of Tan~\emph{et al}.'s QI signal and idler from (\ref{SigFourier}) and (\ref{IdlFourier}) have associated photon annihilation operators, $\hat{a}_{S_m}$ and $\hat{a}_{I_m}$, whose adjoints are the photon creation operators, $\hat{a}^\dagger_{S_m}$ and $\hat{a}^\dagger_{I_m}$.  These names originate from the operators' actions on their mode's number states, which, for $K = S,I$, obey
\begin{equation}
\hat{a}_{K_m}|n\rangle_{K_m} = \left\{\begin{array}{ll}\sqrt{n}\,|n-1\rangle_{K_m}, & \mbox{for $n = 1,2,\ldots,$}\\[.05in] 0,&\mbox{for $n=0$,}\end{array}\right.
\end{equation}
and
\begin{equation}
\hat{a}^\dagger_{K_m}|n\rangle_{K_m} = \sqrt{n+1}\,|n+1\rangle_{K_m}, \mbox{ for $n=0,1,2,\ldots,$}
\end{equation}
where the ket vectors $\{|n\rangle_{K_m} : n=0,1,2,\ldots,\}$ represent states containing exactly $n$ photons.  Ideal photon counting on the $m$th signal and idler modes measures the photon-number operators $\hat{N}_{S_m} \equiv \hat{a}^\dagger_{S_m}\hat{a}_{S_m}$ and 
$\hat{N}_{I_m} \equiv \hat{a}^\dagger_{I_m}\hat{a}_{I_m}$, respectively.  It then follows that ideal photon counting on the coherent state given in~(\ref{CSnumberrep}) yields a Poisson-distributed output with mean $N_S$, as expected from the coherent state's photodetection statistics being obtainable from semiclassical (shot-noise) theory~\cite{Shapiro2009}.  

The signal and idler's $m$th temporal modes have operator-valued quadrature components,
\begin{equation}
{\rm Re}(\hat{a}_{K_m}) = \frac{\hat{a}_{K_m}+\hat{a}^\dagger_{K_m}}{2} \quad\mbox{and}\quad
{\rm Im}(\hat{a}_{K_m}) = \frac{\hat{a}_{K_m}-\hat{a}^\dagger_{K_m}}{2j},
\label{quadratures}
\end{equation}
for $K = S,I$, and ideal (quantum-limited) optical homodyne detection with the appropriate local oscillator fields measures these operators~\cite{Shapiro2009}.   The positive-operator-valued measurements (POVMs) associated with $\hat{a}_{S_m}$ and $\hat{a}_{I_m}$~\cite{Weedbrook2012} can be realized by ideal (quantum-limited) optical heterodyne detection~\cite{Shapiro2009}.  Although heterodyne detection provides information about both quadratures, the Heisenberg uncertainty principle forces this measurement to incur extra noise on each quadrature that is not present in homodyne measurement of a single quadrature~\cite{Shapiro2009}.

Gaussian states of the $m$th signal and idler modes, $\hat{a}_{S_m}$ and $\hat{a}_{I_m}$, are the quantum analogs of classical, complex-valued Gaussian random variables, $a_{S_m}$ and $a_{I_m}$.  Thus, Gaussian states of these two modes are completely characterized by knowledge of their first and second moments~\cite{Weedbrook2012}, 
$\langle \hat{a}_{K_m}\rangle$, $\langle \Delta \hat{a}^\dagger_{K_m}\Delta \hat{a}_{J_m}\rangle$, and $\langle \Delta \hat{a}_{K_m}\Delta \hat{a}_{J_m}\rangle$,
where $K= S,I$, $J=S,I$, $\langle \cdot\rangle$ denotes ensemble average, and $\Delta \hat{a}_{K_m} \equiv \hat{a}_{K_m}-\langle \hat{a}_{K_m}\rangle$.  The coherent state from~(\ref{CSnumberrep}) is a Gaussian state with $\langle \hat{a}_{S_m}\rangle = \sqrt{N_S}$, and $\langle \Delta \hat{a}^\dagger_{S_m}\Delta \hat{a}_{S_m}\rangle =  \langle \Delta \hat{a}_{S_m}^2\rangle = 0$.  The entangled signal-idler state from~(\ref{TMSV}) is also Gaussian.  It is a two-mode squeezed vacuum (TMSV) state~\cite{Weedbrook2012}, whose measurement statistics are completely characterized by $\langle \hat{a}_{S_m}\rangle = \langle \hat{a}_{I_m}\rangle = 0$, $\langle \hat{a}^\dagger_{S_m}\hat{a}_{S_m}\rangle = \langle \hat{a}^\dagger_{I_m}\hat{a}_{I_m}\rangle = N_S$, $\langle \hat{a}_{S_m}^2 \rangle = \langle \hat{a}_{I_m}^2\rangle = 0$, $\langle \hat{a}^\dagger_{S_m}\hat{a}_{I_m}\rangle = 0$, and $\langle \hat{a}_{S_m}\hat{a}_{I_m}\rangle = \sqrt{N_S(N_S+1)}$. 

The preceding brief introduction to Gaussian states provides enough information to understand the origin of Tan~\emph{et al}.'s QI advantage \emph{and} why conventional optical receivers---direct detection, homodyne detection, and heterodyne detection---do not realize any of that advantage.   The cross correlations of all zero-mean \emph{classical} signal-idler states must obey~\cite{footnote2}
\begin{equation}
|\langle \hat{a}^\dagger_{S_m}\hat{a}_{I_m}\rangle| \le \sqrt{\langle \hat{a}^\dagger_{S_m}\hat{a}_{S_m}\rangle\langle \hat{a}^\dagger_{I_m}\hat{a}_{I_m}\rangle},
\label{PhaseInsensitive}
\end{equation}
and
\begin{equation}
|\langle \hat{a}_{S_m}\hat{a}_{I_m}\rangle| \le \sqrt{\langle \hat{a}^\dagger_{S_m}\hat{a}_{S_m}\rangle\langle \hat{a}^\dagger_{I_m}\hat{a}_{I_m}\rangle}.
\label{PhaseSensitive}
\end{equation}
For arbitrary, zero-mean quantum states (\ref{PhaseInsensitive}) applies, but (\ref{PhaseSensitive}) becomes the less restrictive condition~\cite{footnote3}
\begin{equation}
|\langle \hat{a}_{S_m}\hat{a}_{I_m}\rangle| \le \sqrt{\max_{K=S,I}(\langle \hat{a}^\dagger_{K_m}\hat{a}_{K_m}\rangle)\min_{K=S,I}(\langle \hat{a}^\dagger_{K_m}\hat{a}_{K_m}\rangle+1)}.
\label{PhaseSensitiveQ}
\end{equation}

The TMSV is a zero-mean Gaussian state that violates (\ref{PhaseSensitive}), because $|\langle \hat{a}_{S_m}\hat{a}_{I_m}\rangle| = \sqrt{N_S(N_S+1)} > N_S = \sqrt{\langle \hat{a}^\dagger_{S_m}\hat{a}_{S_m}\rangle\langle \hat{a}^\dagger_{I_m}\hat{a}_{I_m}\rangle}$.  Thus it is a nonclassical state, as we already knew from~(\ref{TMSV}), and in fact \emph{maximally} entangled, because it saturates the bound in (\ref{PhaseSensitiveQ}).  Furthermore, with $\hat{a}_{R_m} = \sqrt{\kappa}\,\hat{a}_{S_m} + \sqrt{1-\kappa}\,\hat{a}_{B_m}$ being the photon annihilation operator for the returned light's $m$th mode when the target is present, where $\hat{a}_{B_m}$ is the photon annihilation operator of the relevant background-light mode~\cite{Tan}, Tan~\emph{et al}.'s QI system has the conditional cross-correlation
\begin{equation}
\langle \hat{a}_{R_m}\hat{a}_{I_m}\rangle_{H_1} = \sqrt{\kappa N_S(N_S+1)},
\end{equation} 
given the target-present hypothesis $H_1$.  When $N_S \ll 1$, this conditional cross-correlation, which is the signature of target presence in Tan~\emph{et al}.'s QI system because $\langle \hat{a}_{R_m}\hat{a}_{I_m}\rangle_{H_0} = 0$, greatly exceeds the $\sqrt{\kappa}\, N_S$ classical-state limit on this cross correlation that applies when $\langle \hat{a}^\dagger_{S_m}\hat{a}_{S_m}\rangle = \langle \hat{a}^\dagger_{I_m}\hat{a}_{I_m}\rangle = N_S$.  When $N_S \gg 1$, however, QI's target-presence signature is only slightly better than the classical limit~\cite{footnote4}.  So, it should be no surprise that the preferred operating regime for Tan~\emph{et al.}'s QI system has low signal brightness, $N_S \ll 1$.  That this preferred operating regime has high background brightness, $N_B \gg 1$, follows from Nair's no-go theorem~\cite{Nair2011}, which shows that QI offers only an inconsequential performance improvement over CS operation's $\Pr(e)_{\rm CS} \le e^{-M\kappa N_S}/2$ when background light can be neglected.  

It might seem it would be easy to build a receiver capable of reaping the benefit associated with QI's enhanced cross-correlation signature in the $0<\kappa \ll 1$, $N_S\ll 1$, $N_B\gg 1$ regime.  After all, $\hat{a}_{R_m}\hat{a}_{I_m}$ can be expanded into four terms that are products of the return and idler modes' quadrature components, and these quadrature components can be measured individually by means of homodyne detection.  Unfortunately, this approach does not work, because we need to measure \emph{both} quadratures of the return and idler modes, and the Heisenberg uncertainty principle precludes that being done, e.g., by heterodyne detection, without incurring additional noise.  The net effect is that neither homodyne detection nor heterodyne detection can be used to provide any QI target-detection performance advantage over its CS counterpart.  Direct detection (time-resolved photon counting) is also unable to provide a QI advantage.  Although SPDC produces signal and idler photons in time-coincident photon pairs, photon-coincidence counting---as routinely used, e.g., in SPDC ghost imaging~\cite{Shapiro2012}---does not provide a usable QI signature for target presence.  This failure is because the presence of high-brightness background light in the return implies that every detection of an idler photon will have a coincident detection from the returned light, regardless of target absence or presence.

The signature of target presence in Tan~\emph{et al}.'s QI system is the conditional \emph{phase-sensitive} cross correlation between its returned light and its retained idler, namely $\langle \hat{a}_{R_m}\hat{a}_{I_m}\rangle_{H_1}$ for those beams' $m$th modes.  The returned and retained light's target-present \emph{phase-insensitive} cross correlation, $\langle \hat{a}^\dagger_{R_m}\hat{a}_{I_m}\rangle_{H_1}$, can be measured in second-order interference between the two beams.  But Tan~\emph{et al}.'s QI system has $\langle \hat{a}^\dagger_{R_m}\hat{a}_{I_m}\rangle_{H_j} = 0$ for $j=0,1$, i.e., regardless of whether the target is absent or present, and its $\langle \hat{a}_{R_m}\hat{a}_{I_m}\rangle_{H_j}$ cannot be measured in second-order interference; see Appendix~\ref{AppC} for the details.  Guha and Erkmen~\cite{GuhaErkmen} recognized this problem, and offered a partial solution with their optical parametric amplifier (OPA) receiver.  That receiver, its experimental realization, and other aspects of obtaining QI's performance advantage are treated in the next section.

\section*{QI Receivers and Experiments}
To understand Guha and Erkmen's OPA receiver requires some results from the quantum theory of nonlinear optics.  Because that theory also underlies the SPDC behavior we have already been employing, a brief introduction encompassing both SPDC and OPA operation is germane.

In crystals that have a  second-order nonlinear susceptibility, such as lithium niobate (LiNbO$_3$) or potassium titanyl phosphate (KTiOPO$_4$), a strong pump beam at frequency $\omega_P$ can interact with weak signal and idler beams at lower frequencies $\omega_S$ and $\omega_I$ satisfying $\omega_S +\omega_I = \omega_P$.  Continuous-wave SPDC has no inputs at the signal and idler frequencies.  Nevertheless, it produces outputs at those frequencies. These outputs can be regarded as arising from a photon-fission process, in which a single pump photon splits into a signal-idler photon pair.  Energy conservation at the single-photon level requires that $\hbar\omega_S+\hbar\omega_I = \hbar\omega_P$.   Momentum conservation at the single-photon level requires that $\hbar\bk_S + \hbar\bk_I = \hbar\bk_P$, where $\bk_J$, for $J = S,I,P$, are the propagation vectors for the signal, idler, and pump photons.  Even with these conservation conditions obeyed---which can only be achieved within the crystal's phase-matching bandwidth, $|\omega_K -\omega_{K_o}| \le \pi W$ for $K=S,I$, about the signal and idler's center frequencies, $\omega_{S_o}$ and $\omega_{I_o}$---continuous-wave SPDC produces only pW of signal and idler per mW of pump power, see, e.g.,~\cite{Zhong2012}.  

For the QI application, we want the signal and idler to be single-spatial-mode fields, so the former can be formed into a tight transmitter beam, which is why our description of Tan~\emph{et al}.'s QI system---and, similarly, of Lloyd's QI system---only considered temporal modes.  The Gaussian-state treatment of single-spatial-mode SPDC~\cite{Wong2006} shows that the signal's frequency component at $\omega_{S_o} + \omega$ is only correlated with the idler's frequency component at $\omega_{I_o} - \omega$, which explains Appendix~\ref{AppB}'s choice of temporal modes, and justifies our only paying attention to correlations between the $m$th modes of the signal and idler in our treatment of Tan~\emph{et al}.'s QI system.  

Guha and Erkmen's receiver for Tan~\emph{et al}.'s QI system uses a continuous-wave OPA operating at extremely low gain, which means that it can be realized with a crystal identical to the one used for that system's SPDC source.  In the receiver, however, the returned light and the retained idler light are applied as the signal and idler inputs to the OPA crystal, and the idler output's $m$th mode is given by $\hat{a}_{I_m}^{\rm out} = \sqrt{G}\,\hat{a}_{I_m} + \sqrt{G-1}\,\hat{a}_{R_m}^\dagger$, where the gain, $G>1$, is chosen to optimize target-detection performance.  Ideal photon counting is done on all idler-output modes, which measures their total photon-number operator, 
\begin{equation}
\hat{N}_T \equiv \sum_{m=-(M-1)/2}^{(M-1)/2}\hat{a}_{I_m}^{{\rm out}\dagger}\hat{a}_{I_m}^{\rm out}.
\end{equation}
Although written as a sum of modal photon-number operators, the $\hat{N}_T$ measurement is easily realized by photon counting on the OPA's duration-$T$ idler output in response to the duration-$T$ returned light and retained idler inputs.  The conditional means---given target absence or presence---for the $\hat{N}_T$ measurement are as follows:
\begin{align}
\langle\hat{N}_T\rangle_{H_j} \!= &\!\!\!\sum_{m=-(M-1)/2}^{(M-1)/2}\left(G\langle\hat{a}_{I_m}^\dagger\hat{a}_{I_m}\rangle + (G-1)\langle \hat{a}_{R_m}\hat{a}_{R_m}^\dagger\rangle_{H_j}\right. \nonumber \\[.05in] 
&+ \left. 2\sqrt{G(G-1)}\,{\rm Re}[\langle\hat{a}_{R_m}\hat{a}_{I_m}\rangle_{H_j}] \right).
\end{align}
Here, the phase conjugation of the returned light's contribution to the OPA's idler output makes QI's phase-sensitive cross-correlation signature of target presence observable in a photon-counting measurement, cf.~Appendix~\ref{AppC}, where without phase conjugation it is only the phase-insensitive cross correlation that can be measured in a photon-counting interferometer.  

It is the high brightness of the returned light's background component, in comparison the low brightness of the retained idler, that leads to the OPA receiver's optimum gain satisfying $G-1 \ll 1$.  At this optimum gain value, Guha and Erkmen found their receiver's quantum Chernoff bound to be
\begin{equation}
\Pr(e)_{\rm OPA} \le e^{-M\kappa N_S/2N_B}/2,
\label{OPAperf}
\end{equation}
in QI's usual $0<\kappa\ll 1$, $N_S \ll 1$, $N_B \gg 1$, operating regime, implying a 3\,dB improvement in error-probability exponent over Tan~\emph{et al.}'s CS system of the same transmitted energy.  

Implicit in (\ref{OPAperf}) are two important assumptions made by Guha and Erkmen, both of which arise from the interferometric nature of applying the returned light and the retained idler to the signal and idler inputs of a low-gain OPA.  The first assumption is that the idler is stored losslessly.  The second is that the idler storage is matched in time delay and phase to those of the light returned from the target (when it is present).  We will revisit these assumptions later in discussing QI's utility for realistic radar scenarios.  For now, it suffices to note that Zhang~\emph{et al}.~\cite{Zhang} reported the first experimental demonstration of Tan~\emph{et al}.'s QI system using an OPA receiver in a table-top setup.  Owing to a variety of experimental nonidealities, that experiment yielded only a 20\% signal-to-noise ratio (SNR) improvement---equivalent to a 20\% error-probability exponent advantage---over a CS system of the same transmitted energy.

Prior to Zhang~\emph{et al}.'s work, Lopaeva~\emph{et al}.~\cite{Lopaeva} reported a QI-like experiment in which an SPDC source and a photon-counting correlations were used to obtain an SNR advantage over a correlated-thermal-state (correlated-noise radar) probe.  That experiment, however, did not exploit entanglement, and its QI-like system only outperformed a correlated-thermal-state system of the same transmitted energy, not a CS system of that energy.  A more QI recent experiment, by England~\emph{et al}.~\cite{England}, used an SPDC source and photon-coincidence counting for QI, but it operated in the low-brightness, $N_B\ll 1$, background regime, wherein QI offers no advantage over a coherent-state radar.

The failure of OPA reception to achieve QI's full 6\,dB advantage in error-probability exponent is a consequence of Tan~\emph{et al}.'s QI scenario being one of mixed-state hypothesis testing and OPA reception's using mode-pair measurements.  Such an arrangement falls into the class of local operations plus classical communication (LOCC) processing, which is known~\cite{Calsamiglia2010}, \cite{Bandyopadhyay2011} to be suboptimal for mixed-state hypothesis testing.
There is an in-principle approach to realizing QI's full performance advantage over CS operation---a Schur transform on a quantum computer~\cite{Bacon}---but Zhuang~\emph{et al}.~\cite{Zhuang} have 
proposed a receiver structure for this purpose that does not require a full-blown quantum computer.  They did so by  exploiting SPDC's inverse operation, sum-frequency generation, to go beyond the bounds set by LOCC processing. 

A signal-idler photon pair produced by continuous-wave SPDC with phase-matching bandwidth $W$ is in a quantum state satisfying
\begin{equation}
|\psi\rangle_{SI} \propto \int_{-\pi W}^{\pi W}\!\frac{{\rm d}\omega}{2\pi}\,|\omega_{S_o}+\omega\rangle_S|\omega_{I_o}-\omega\rangle_I,
\end{equation}
where $|\omega_{S_o}+\omega\rangle_S$ ($|\omega_{I_o}-\omega\rangle_I$) denotes a single-photon signal (idler) of frequency $\omega_{S_o}+\omega$ ($\omega_{I_o}-\omega$) and, as has been implicitly assumed earlier, the signal (idler) brightness is taken to be constant over the phase-matching bandwidth.  If this photon pair illuminates a crystal identical to the one used for its SPDC generation, then, with \emph{extremely} low probability, sum-frequency generation (SFG) can occur, viz., a photon-fusion process in which the signal-idler photon pair is converted to a single photon at the pump frequency, $\omega_P = \omega_{S_o}+\omega_{I_o}$.   Zhuang~\emph{et al}.~realized that SFG's being a coherent process involving all of QI's mode pairs offered a path to QI reception that was not bound by the limitations of LOCC operation.  Nevertheless, their work made the rather significant assumption that SFG could be done with 100\% efficiency at the photon-pair level, something that is far beyond the current capability of nonlinear optics.  Furthermore, the presence of high-brightness background light drove them to using multiple cycles of SFG and photon-counting measurements, but ideal realization of these steps resulted in a receiver whose quantum Chernoff bound for QI matched that found by Tan~\emph{et al}.  Zhuang~\emph{et al}.~augmented their SFG receiver with an appropriate feed-forward (FF) circuit---inspired by the Dolinar receiver~\cite{Dolinar} for optimum quantum reception of coherent-state signals---and showed that the resulting FF-SFG receiver provided minimum error-probability performance in QI target detection.  Despite this receiver's being well beyond the reach of available technology, it is nevertheless important because it allowed determination of QI's receiver operating characteristic (ROC), i.e., its detection probability, $P_D$, versus false-alarm probability, $P_F$, when optimum quantum reception is employed~\cite{ZhuangROC}.  QI's ROC is crucial because it is a far better target-detection performance metric than error probability, as radar targets should not be presumed equally likely to be absent or present.  Not surprisingly, QI's ROC improvement over CS operation of the same transmitted energy turned out to be equivalent to a 6\,dB increase in effective SNR~\cite{ZhuangROC}.

QI's aforementioned performance advantages---in error probability or ROC---assume that the target return, when present, has known amplitude and phase, a situation that seldom occurs in light detection and ranging (lidar) applications. At lidar wavelengths,
most target surfaces are sufficiently rough that their returns are speckled, i.e., they have Rayleigh-distributed amplitudes and uniformly-distributed phases. QI's OPA receiver---which affords a 3\,dB-better-than-classical error-probability exponent for a return with known amplitude and phase---fails to offer any performance gain for Rayleigh-fading targets. The SFG receiver from~Ref.~\cite{Zhuang}---whose error-probability exponent for a nonfading target achieves QI's full 6\,dB advantage over optimum classical operation---outperforms the classical system for Rayleigh-fading targets~\cite{ZhuangRayleigh}. In this case, however, QI's advantage is subexponential under ideal operating conditions, so that its benefit is far more vulnerable to nonidealities such as were encountered in Zhang~\emph{et al}.'s OPA receiver for the non-fading scenario.

\section*{Microwave Quantum Illumination:  Concept and Realities}
Lloyd's QI presumed operation at optical wavelengths, because high-sensitivity photodetection systems have long been limited by noise of quantum-mechanical origin.  Hence the optical region was the natural setting in which to seek a quantum advantage.  Tan~\emph{et al}.~continued to focus on optical wavelengths, because that is where SPDC sources provide the entanglement needed for Gaussian-state QI.  Unfortunately, although Tan~\emph{et al}.~found QI's performance to exceed that of all classical radars of the same transmitted energy, the regime in which that occurred, $0<\kappa\ll 1$, $N_S \ll 1$, and $N_B \gg 1$, is not realistic for optical wavelengths.  Weakly-reflecting (low radar cross-section) targets are of great interest, and SPDC sources naturally produce $N_S \ll 1$ emissions, but $N_B \gg 1$ does not prevail at optical wavelengths.  For example, a typical value for the sky's daytime spectral radiance at the eyesafe 1.55\,$\mu$m wavelength is $N_\lambda \sim 10$\,W/m$^2$\,SR\,$\mu$m \cite{Kopeika1970}, from which we find that~\cite{Shapiro2005}
\begin{equation}
N_B = \pi 10^6\lambda^3N_\lambda/\hbar\omega^2 \sim 10^{-6},
\end{equation}
and nighttime $N_\lambda$ (and hence $N_B$) values are several orders of magnitude lower.  So, the advantage expected from ideal QI operation at optical frequencies would only accrue were there bright-light jamming.  Consequently, a significant amount of interest in QI from the radar community only arose after Barzanjeh~\emph{et al}.~\cite{Barzanjeh} proposed an approach for doing QI in the microwave region, where the naturally-occurring background does satisfy $N_B \gg 1$~\cite{footnote5}, and most target-detection radars operate.  

\begin{figure*}[hbt]
\centering
\includegraphics[width=6in]{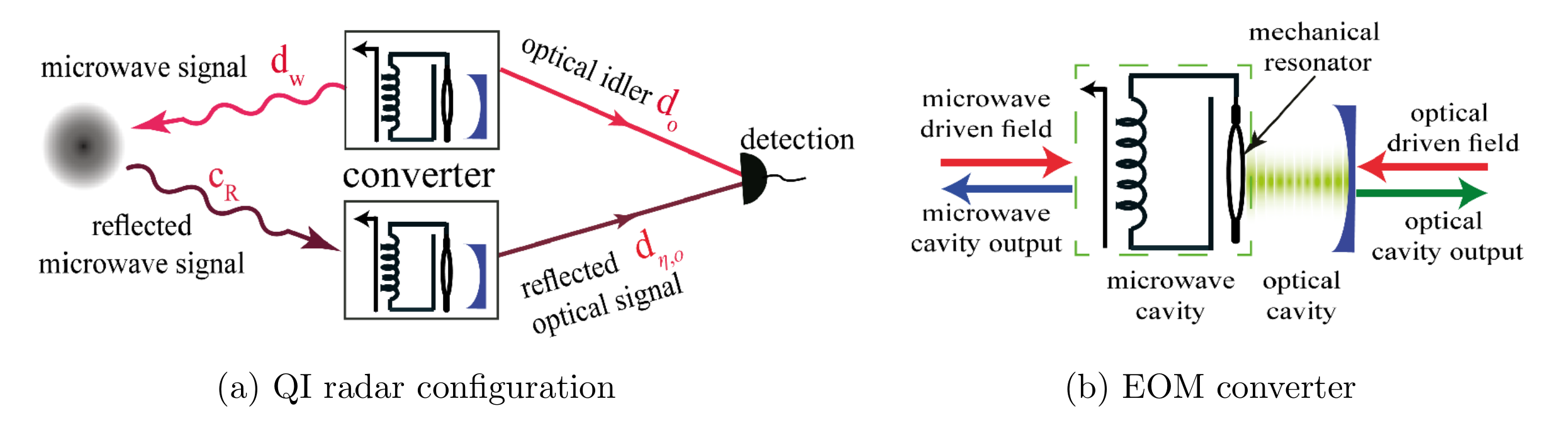}
\caption{Schematic for Barzanjeh~\emph{et al.}'s microwave quantum illumination~\cite{Barzanjeh}.  (a) QI radar configuration.  (b) Electro-optomechanical (EOM) converter configuration.  
\label{QI_fig4}}
\end{figure*}

Figure~\ref{QI_fig4} shows Barzanjeh~\emph{et al}.'s transmitter and receiver concepts.  Their transmitter uses an electro-optomechanical (EOM) converter to create a microwave signal that is entangled with an optical idler, and it transmits the microwave signal to irradiate the region in which the target may be present.  Then, it uses another EOM converter to upconvert the returned microwave signal to the optical region for a phase-conjugate (PC) joint measurement with the retained idler.  Guha and Erkmen~\cite{GuhaErkmen} had previously shown that a PC receiver's performance advantage over a conventional radar is equivalent to that of an OPA receiver, i.e., a 3\,dB advantage in error-probability exponent under ideal conditions.  Even though this advantage could easily fall prey to system nonidealities, the fact that QI was now predicted to offer an advantage at microwave frequencies ignited a great deal of attention from the radar community, prompted in part by some inaccurate reporting~\cite{ExtremeTech2015}.  It is now time, therefore, to turn our attention to those nonidealities, some of which were discussed in Refs.~\cite{Barzanjeh}, \cite{Pirandola2018}, and confront the realities of seeking a QI advantage for target detection.  We will begin that assessment with the issue of idler-storage loss.

The OPA, PC, SFG, and FF-SFG receivers for QI all require that the idler be stored for a roundtrip, radar-to-target-to-radar, propagation time.  If the idler-storage transmissivity is $\kappa_I$ ($0 < \kappa_I \le 1$), then the quantum Chernoff bounds of interest for non-fading targets become
\begin{equation}
\Pr(e)_{\rm QI} \le 
e^{-M\kappa\kappa_IN_S/N_B}/2, 
\end{equation}
for FF-SFG or SFG receivers,
and
\begin{equation}
\Pr(e)_{\rm QI} \le 
e^{-M\kappa\kappa_IN_S/2N_B}/2, 
\end{equation}
for OPA or PC receivers,
in QI's preferred $0 < \kappa\ll 1$, $N_S\ll 1$, $N_B \gg 1$ operating regime.  Comparing these results with (\ref{TanCS})'s quantum Chernoff bound for a CS radar shows that 6\,dB of idler loss will eliminate QI's performance advantage for FF-SFG or SFG reception, and 3\,dB of idler loss suffices for that purpose for OPA or PC reception.  Barzanjeh~\emph{et al.}~\cite{Barzanjeh} was the first QI target-detection paper to explicitly comment on the adverse impact of idler-storage loss, noting that the idler-storage loss incurred in the best presently-available technique for optical idler storage---an optical-fiber delay line---precludes their system's offering any performance advantage for targets more than 11.25\,km away.  

Even were the preceding gloomy assessment of long-range, microwave-QI target detection obviated by a breakthrough in idler-storage technology, microwave QI would still suffer from its difficulty in obtaining a sufficiently high time-bandwidth product with pulse durations that make sense for microwave radars.  High time-bandwidth product is critical for QI because---unlike a conventional radar, whose performance  improves at constant pulse duration and bandwidth with increasing transmitter power in the absence of clutter---QI systems' performance advantage degrades at constant pulse duration and bandwidth with increasing transmitter power~\cite{Pirandola2018}.  So, even though Barzanjeh~\emph{et al}.'s EOM converters---which are intrinsically narrowband---could be replaced with broadband microwave-entanglement generation in a traveling-wave parametric amplifier~\cite{Oliver}, the available time-bandwidth products in the microwave region are still dwarfed by what is easily achieved at optical wavelengths.  For example, the 1/3-percent fractional bandwidth at 1\,$\mu$m wavelength that gives a 1\,$\mu$s pulse duration a $10^6$ time-bandwidth product only gives that pulse duration a $10^2$ time-bandwidth product at 1\,cm wavelength. Pushing to mm-wave operation with higher fractional bandwidth and longer pulse duration will afford considerably higher time-bandwidth product, but that will run afoul of another consideration: QI's  single-bin-per-pulse interrogation limitation~\cite{Pirandola2018}.    

Unlike a conventional radar, QI can only interrogate a single polarization-azimuth-elevation-range-Doppler resolution bin at a time~\cite{footnote6}, if it is derive its full performance advantage over a conventional radar.  Interrogation of $K_B$ bins simultaneously, by splitting the stored idler into $K_B$ equal-strength pieces to make said measurements, leads to a $10\log_{10}(K_B)$\,dB performance loss for each bin.  So, assuming ideal equipment, simultaneous interrogation of two resolution bins sacrifices the entire performance advantage of an OPA receiver, and simultaneous interrogation of four resolution bins gives up the entire performance advantage of an FF-SFG receiver.  In this regard it is important to note that 
optical amplification cannot be used, prior to idler splitting, to mitigate the preceding problem.  This failure is because the amplifier's unavoidable amplified spontaneous emission noise will be much stronger than the amplified idler, making the latter useless for QI.  There is another resolution-related problem with QI:  the target should lie entirely within a single polarization-azimuth-elevation-range-Doppler resolution bin throughout the signal pulse's full duration.  If such is not the case, there will be a mismatch between the temporal behavior of the returned radiation's target-present component and the temporal behavior of the stored idler.  This mismatch, which will degrade the performance of the QI receivers we have considered, is quantified by a normalized overlap integral, $0 < \kappa_m\le 1$, that reduces QI's error-probability exponent by a factor of $\kappa_m$.  The mismatch problem is especially significant for optical QI, where a 1\,THz bandwidth implies that $<$1\,mm target range extent is needed to ensure $\kappa_m \approx 1$.  

Some additional points worth noting are as follows:  (1) Las Heras~\emph{et al}.~\cite{QIcloak} have claimed that QI target detection can unveil an electromagnetically-cloaked target from the phase shift it creates, hence potentially reviving the radar community's interest in QI.  But Las Heras~\emph{et al}.~assume an OPA receiver and neglect idler-storage loss, so their concept's utility is limited by the same considerations cited above for Barzanjeh~\emph{et al}.'s system.  (2) De Palma and Borregaard~\cite{DePalma} have shown that the state produced by SPDC is the optimum transmitter state for Tan~\emph{et al.}'s QI scenario if the objective is to obtain the fastest decay of the miss probability, $P_M \equiv 1- P_D$, for a given false-alarm probability, $P_F$, as the transmitted energy is increased.  In effect, their result implies that the SPDC state optimizes the ROC for Tan~\emph{et al.}'s QI scenario.  So, because QI target detection for equally-likely target absence or presence has an error probability satisfying $\Pr(e) = [P_F + (1-P_D)]/2$, where $(P_F, P_D)$ lies on the system's ROC, De Palma and Borregaard's result provides strong evidence that a 6\,dB advantage for QI's error-probability exponent is the best that can be done relative to a coherent-state transmitter of the same energy in this equally-likely situation.  (3) Initial microwave QI experiments have been performed by Luong~\emph{et al}.~\cite{Luong2018}, \cite{Luong2019}, and by Barzanjeh~\emph{et al}.~\cite{Barzanjeh2019}.  Both used Josephson junction parametric amplifiers to produce entangled signal and idler in the GHz region, and both performed pre-amplified heterodyne detection at the signal and idler frequencies.  Furthermore, both found substantial performance gains for their QI systems over their chosen classical comparison cases.  In Luong~\emph{et al}.'s experiments the comparison case was a classically-correlated-noise radar, whereas in Barzanjeh~\emph{et al}.'s work comparisons were made with both a classically-correlated-noise radar and a coherent-state radar with incoherent post-heterodyne processing.  As the next section will show, their reception techniques preclude their QI experiments from outperforming an optimized classically-correlated-noise (CCN) radar. For that reason, we will refer to their QI setups as quantum-correlated-noise (QCN) radars, as opposed to a true QI radar, viz., the Tan~\emph{et al}.~scenario with an OPA receiver, which outperforms \emph{all} classical radars of the same transmitted energy~\cite{footnote7}.

\section*{Correlated-Noise Radars:  Quantum versus Classical}
Here, for greater generality in our consideraton of correlated-noise radars, we will use
\begin{equation}
\hat{a}_{R_m} = \sqrt{\kappa}\,e^{j\theta}\hat{a}_{S_m} + \sqrt{1-\kappa}\,\hat{a}_{B_m},\mbox{ for $|m| \le (M-1)/2$},
\end{equation}
to model the returned light's $M$ modal annihilation operators when the target is present, where $0< \kappa \ll 1$ and $0\le \theta\le 2\pi$ may be deterministic and known, as in Tan~\emph{et al}.~\cite{Tan} ($\kappa\ll 1$ known and $\theta=0$), or random with a known joint probability distribution, as in Zhuang~\emph{et al}.~\cite{ZhuangROC} ($\sqrt{\kappa}$ and $\theta$ statistically independent, with a Rayleigh-distributed $\sqrt{\kappa}$ having $\langle \kappa\rangle \ll 1$, and a uniformly-distributed $\theta$).   We will also assume that these noise radars use pre-amplified heterodyne detection, with noise figure $N_F \ge 1$, where 1 is the ideal (quantum-limited) value~\cite{footnote8}.  

The QCN radar we shall consider uses an SPDC source and pre-amplified heterodyne detection at the signal and idler frequencies.  Conditioned on target absence (hypothesis $H_0$) or presence (hypothesis $H_1$) \emph{and} the $\kappa$, $\theta$ values, the $M$ mode-pair outputs from the QCN radar's heterodyne detectors are a set of independent, identically-distributed, complex-valued, 2D random column vectors, $\{\ba_m \equiv [\begin{array}{cc} a_{R_m} & a_{I_m}\end{array}]^T : |m| \le (M-1)/2\}$, 
whose quadrature components have zero-mean Gaussian distributions with covariance matrices
\begin{equation}
\bLam^{\rm QCN} _{H_0} =  \frac{G_A}{2}\left[\begin{array}{cc}
(N_B + N_F){\bf I}_2 & {\bf 0}_2 \\[.05in]
{\bf 0}_2 & (N_S + N_F){\bf I}_2 \end{array}\right],
\end{equation}
and
 \begin{equation}
 \bLam^{\rm QCN} _{H_1,\kappa,\theta} = \frac{G_A}{2}\left[\begin{array}{cc}
 (N_R + N_F){\bf I}_2 & {\bf C}_q(\theta) \\[.05in]
{\bf C}_q^T(\theta) & (N_S + N_F){\bf I}_2 
 \end{array}\right].
 \end{equation}
In these equations: $G_A$ is the pre-amplifiers' gain; ${\bf I}_2$ is the $2\times 2$ identity matrix; ${\bf 0}_2$ is the $2\times 2$ matrix of zeros; $N_R \equiv \kappa N_S + N_B$,; and ${\bf C}_q(\theta) = \sqrt{\kappa N_S(N_S+1)}\,{\bf R}_q(\theta)$ with
\begin{equation}
{\bf R}_q(\theta) \equiv \left[\begin{array}{cc}
\cos(\theta) & \sin(\theta) \\[.05 in]
\sin(\theta) & -\cos(\theta) \end{array}\right].
\end{equation}

The CCN radar we will consider uses a high-brightness classical noise source whose output is divided---with an appropriate splitter---into a low-brightness signal and a high-brightness idler.  Like the QCN radar, the CCN radar will employ pre-amplified heterodyne detection of its retained idler and its return from the region interrogated by its signal.  Conditioned on the true hypothesis and the $\kappa$, $\theta$ values, the CCN radar's $M$ mode-pair outputs are also a set of independent, identically-distributed, complex-valued, 2D random column vectors whose quadrature components have zero-mean Gaussian distributions, but now with covariance matrices
\begin{equation}
\bLam^{\rm CCN} _{H_0} = \frac{G_A}{2}\left[\begin{array}{cc}
(N_B + N_F){\bf I}_2 & {\bf 0}_2 \\[.05in]
{\bf 0}_2 & (N_I + N_F){\bf I}_2 \end{array}\right],
\end{equation}
and
\begin{equation}
 \bLam^{\rm CCN} _{H_1,\kappa,\theta} = \frac{G_A}{2}\left[\begin{array}{cc}
 (N_R + N_F){\bf I}_2 & {\bf C}_c(\theta) \\[.05in]
{\bf C}_c^T(\theta) & (N_I + N_F){\bf I}_2 
 \end{array}\right],
 \end{equation}
where ${\bf C}_c(\theta) = \sqrt{\kappa N_SN_I}\,{\bf R}_c(\theta)$ with 
\begin{equation}
{\bf R}_c(\theta) \equiv \left[\begin{array}{cc}
\cos(\theta) & -\sin(\theta) \\[.05 in]
\sin(\theta) & \cos(\theta) \end{array}\right].
\end{equation} 

Note that SPDC forces the signal and idler brightnesses at the QCN radar's transmitter to be identical, whereas, by starting with a high-brightness classical noise source and using a highly-asymmetric splitting ratio, the CCN radar's transmitter can have a low-brightness ($N_S\ll 1$) signal that matches that of the QCN radar while retaining a high-brightness ($N_I \gg 1$) idler.  As we will soon see, operating the CCN radar with $N_I \gg 1 \gg N_S$ is a crucial point that was missed in the microwave QI experiments to date~\cite{Luong2018}--\cite{Barzanjeh2019}.  

The relative target-detection performance of the QCN and CCN radars is most conveniently understood by transforming their mode-pair data to $\{\ba'_m \equiv [\begin{array}{cc} a_{R_m} & a^*_{I_m}/\sqrt{N_S+1}\end{array}]^T/\sqrt{G_A}\}$ for the QCN radar, and $\{\ba'_m \equiv [\begin{array}{cc} a_{R_m} & a_{I_m}/\sqrt{N_I}\end{array}]^T/\sqrt{G_A}\}$ for the CCN radar.
Given the true hypothesis and the $\kappa$, $\theta$ values, the QCN radar's transformed mode pairs are independent and identically distributed with zero-mean Gaussian-distributed quadrature components having covariance matrices
\begin{equation}
\bLam^{'\rm QCN} _{H_0} =  \frac{1}{2}\left[\begin{array}{cc}
(N_B + N_F){\bf I}_2 & {\bf 0}_2 \\[.05in]
{\bf 0}_2 & \left(1+\frac{\displaystyle N_F-1}{\displaystyle N_S+1}\right)\!{\bf I}_2 \end{array}\right],
 \end{equation}
 and
 \begin{equation}
 \bLam^{'\rm QCN} _{H_1,\kappa,\theta} = \frac{1}{2}\left[\begin{array}{cc}
(N_R+N_F){\bf I}_2 & \sqrt{\kappa N_S}\,{\bf R}_c(\theta) \\[.05in]
 \sqrt{\kappa N_S}\,{\bf R}_c^T(\theta) & \left(1+ \frac{\displaystyle N_F-1}{\displaystyle N_S+1}\right)\!{\bf I}_2
 \end{array}\right].
 \end{equation}
 Similarly, given the true hypothesis and the $\kappa$, $\theta$ values, the CCN radar's transformed mode pairs are independent and identically distributed with zero-mean Gaussian-distributed quadrature components having covariance matrices
\begin{equation}
\bLam^{'\rm CCN} _{H_0} =  \frac{1}{2}\left[\begin{array}{cc}
(N_B + N_F){\bf I}_2 & {\bf 0}_2 \\[.05in]
{\bf 0}_2 & (1+N_F/N_I){\bf I}_2 \end{array}\right],
 \end{equation}
 and
 \begin{equation}
 \bLam^{'\rm CCN} _{H_1,\kappa,\theta} = \frac{1}{2}\left[\begin{array}{cc}
(N_R+N_F){\bf I}_2 & \sqrt{\kappa N_S}\,{\bf R}_c(\theta) \\[.05in]
 \sqrt{\kappa N_S}\,{\bf R}_c^T(\theta) & (1+N_F/N_I){\bf I}_2 
 \end{array}\right].
 \end{equation}

It is now clear that the QCN and CCN radars have \emph{identical} conditional statistics in the limit $N_I \rightarrow \infty$ when their detectors are quantum limited ($N_F=1$).   Hence we expect that their quantum-limited performance will be virtually the same for $N_I \gg 1$.  Moreover, for non-ideal detectors ($N_F > 1$), the CCN radar \emph{outperforms} the QCN radar when $N_I > N_F(N_S+1)/(N_F-1)$, because the CCN radar's transformed idler modes then have lower noise than those of the QCN radar, while all the other second moments of the two radars are identical.  Furthermore, these conclusions apply for $\kappa$, $\theta$ deterministic and known, as well as for $\kappa$, $\theta$ random with a known joint probability distribution.  Thus we have proven that the QCN radar \emph{cannot} outperform all classical radars of the same transmitted energy and identical detector capabilities.

\section*{ROC Comparisons}
As a capstone for all that has been presented, this section compares the ROCs for five radars in the best possible operating scenario:  quantum-limited detection, $\theta = 0$, and $\kappa$ deterministic and known.   All of these radars are trying to detect the presence of a $\kappa = 0.01$ roundtrip-transmissivity target that is embedded in an $N_B = 20$ background by irradiating the region of interest with $N_{\rm tot} = 2 \times 10^4$ photons on average.  The first two are the QCN and CCN radars from the preceding section, with $N_S = 0.01$, $M = 2\times 10^6$, $N_I = 10^3$, and $N_F =1$.  The third (QI-OPA) uses the QCN radar's transmitter plus OPA reception with gain value $G = 1+N_S/\sqrt{N_B}$~\cite{GuhaErkmen} and an ideal photon counter.  The fourth (CS-Het) and fifth (CS-Hom) radars each transmit a coherent-state pulse of average photon number $N_{\rm tot}$, and perform quantum-limited heterodyne detection and quantum-limited homodyne detection, respectively.  

\begin{figure}
\centering
\includegraphics[width=2.5in]{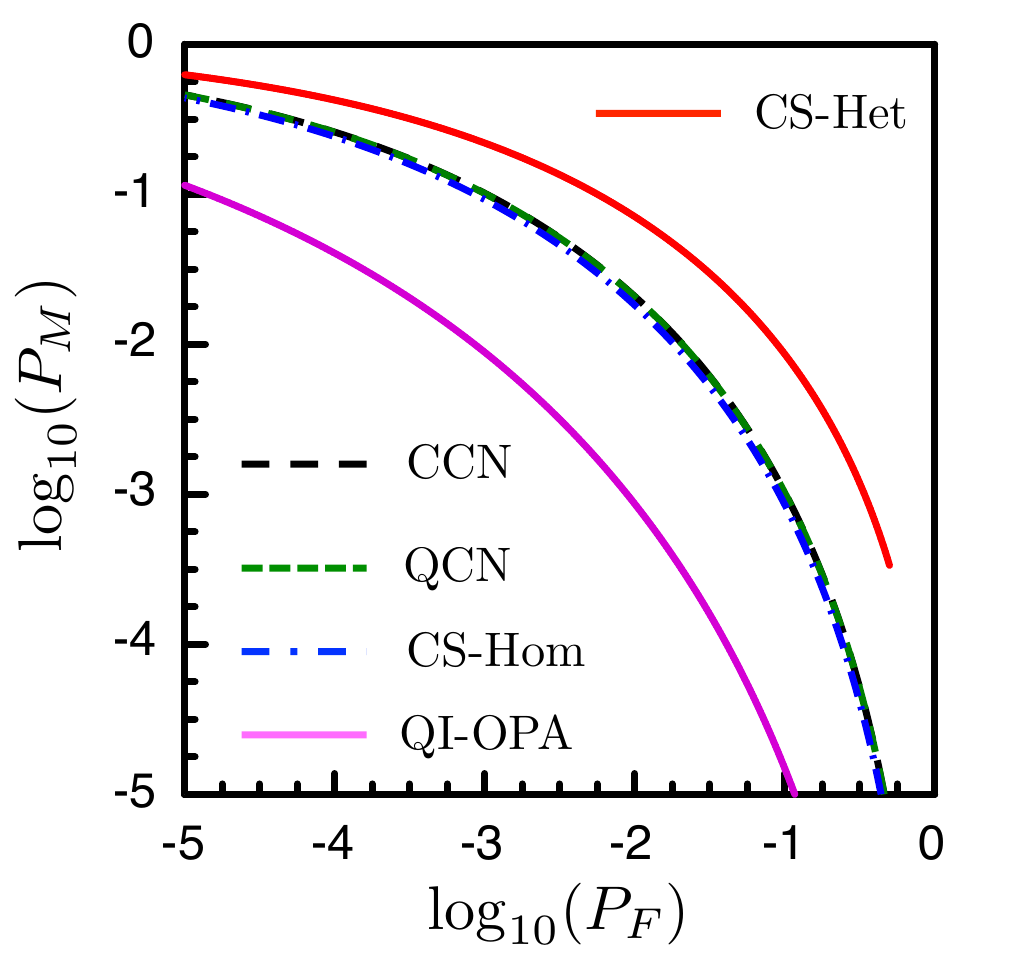}
\caption{ROCs, from top to bottom for the CS-Het, CCN, QCN, CS-Hom, and QI-OPA radars, plotted as $\log_{10}(P_M)$ versus $\log_{10}(P_F)$ for $P_F, P_M \le 0.5$.  All are trying to detect the presence of a $\kappa = 0.01$ roundtrip-transmissivity target that is embedded in an $N_B = 20$ background by irradiating the region of interest with $N_{\rm tot} = 2 \times 10^4$ photons on average.  The QCN and CCN radars have $N_S = 0.01$, $M = 2\times 10^6$, $N_I = 10^3$, and $N_F = 1$.  The QI-OPA radar has $N_S = 0.01$, $M = 2\times 10^6$, OPA gain $G = 1+N_S/\sqrt{N_B}$, and an ideal photon counter.  The CS-Het and CS-Hom radars each transmit a single, coherent-state pulse of average photon number $N_{\rm tot} = 2\times 10^4$, and perform quantum-limited heterodyne detection and quantum-limited homodyne detection, respectively.  
\label{QI_fig5}}
\end{figure}

Figure~\ref{QI_fig5} shows the ROCs for these radars, plotted as $\log_{10}(P_M)$ versus $\log_{10}(P_F)$ for $P_F, P_M \le 0.5$.  The QCN, CCN, and QI-OPA ROCs were computed with Van Trees's ROC approximation technique~\cite{VanTrees}, which is known to be accurate for $M \gg 1$ and especially so~\cite{Shapiro1999} for Gaussian signals in Gaussian noise, as we have for the QCN and CCN radars.  The CS-Het ROC~\cite{VanTrees},
\begin{equation}
P_D = Q(\sqrt{2M\kappa N_S/(N_B+1)},\sqrt{-2\ln(P_F)}),
\end{equation}
where 
\begin{equation}
Q(\alpha,\beta) \equiv \int_{\beta}^\infty\,{\rm d}u\,ue^{-(u^2+\alpha^2)/2}I_0(u\alpha),
\end{equation}
with $I_0(x)$ being the zeroth-order modified Bessel function of the first kind, was computed exactly, as was 
the CS-Hom ROC~\cite{VanTrees},
\begin{equation}
P_D = Q(Q^{-1}(P_F)-\sqrt{4M\kappa N_S/(2N_B+1)}),
\end{equation}
where
\begin{equation}
Q(x) \equiv \int_x^\infty\!{\rm d}u\,\frac{e^{-u^2/2}}{\sqrt{2\pi}},
\end{equation}
with $Q^{-1}(\cdot)$ being its inverse function, which obeys $Q^{-1}[Q(x)] = x$ for $-\infty < x < \infty$.   
 
Figure~\ref{QI_fig5} shows two expected behaviors:  (1) the QCN and CCN ROCs are indistinguishable; and (2) the QI-OPA radar outperforms the CS-Hom, CCN, and CS-Het radars.  It also shows one unexpected behavior:  the CS-Hom and CCN ROCs are also indistinguishable, see Appendix~\ref{AppD} for an explanation of why this occurs. Note that the QCN, CCN, CS-Hom, and QI-OPA radars all use phase information, whereas the CS-Het radar does not, because its receiver performs incoherent (envelope) detecton at the intermediate frequency.  For the QCN and CCN radars with known $\theta$, phase information is used to perform coherent detection at their receivers' intermediate frequency.  For uniformly-distributed $\theta$, the QCN and CCN radars could use incoherent detection at their heterodyne receivers' intermediate frequency, and suffer ROC degradation similar to the performance lost in going from the CS-Hom ROC to the CS-Het ROC.  On the other hand, the CS-Hom and QI-OPA radars \emph{require} phase information: the CS-Hom radar needs that information to lock its local oscillator's phase to that of the expected target return, while the QI-OPA radar needs it to lock its retained idler's phase to that of the expected target return.  One final comment about QCN versus CCN radars is in order:  the QCN radar requires a dilution refrigerator to house its Josephson junction parametric amplifier source, but the CCN radar might use a much less expensive room-temperature microwave noise generator for its source.    

\section*{Concluding Remarks} 
In conclusion, despite QI target detection's not being an immediate boon to the radar community, it is important to remember that it is the first example of a bosonic system that offers a performance improvement on an entanglement-breaking channel when its transmitter is subject to an energy constraint~\cite{footnote9}.  In other words, we should \emph{not} dismiss the value of entanglement for the lossy, noisy situations that are the norm in microwave radar.  Indeed, this meta-lesson may lead to other quantum systems that offer real utility.  Two such examples have already arisen in secure communication:  (1) floodlight quantum key distribution (FL-QKD)~\cite{ZhuangFloodlight}, \cite{ZhangFloodlight}, which directly descends from a communication version~\cite{Shapiro} of QI target detection, is capable of Gbps secret-key rates over metropolitan-area fiber connections; and (2) quantum low probability of intercept~\cite{QLPI}---a repurposed version of FL-QKD---provides security for Gbps ciphertext transmissions over metropolitan-area fiber links despite the presence of an eavesdropper who holds the decryption key.

\section*{Acknowledgments}Preparation of this article was supported by the Air Force Office of
Scientific Research (AFOSR) MURI program under Grant No.~FA9550-14-1-0052 and by the MITRE Corporation's Quantum Moonshot program.  Portions of this paper were previously presented at CLEO 2017~\cite{CLEO2017} and COMCAS 2019~\cite{COMCAS2019}.  The author acknowledges the contributions made to his understanding of quantum illumination by his collaborators (in alphabetical order):  S. Barzanjeh, B. I. Erkmen, V. Giovannetti, S. Guha, S. Lloyd, L. Maccone, S. Mouradian, S. Pirandola, S.-H. Tan, D. Vitali, C. Weedbrook, F. N. C. Wong, Z. Zhang, and Q. Zhuang.  He also acknowledges valuable discussions with and encouragement from G. Gilbert.

\appendix
 \section{Entangled State for Lloyd's QI System \label{AppA}}
 For the high time-bandwidth product ($M=TW \gg 1$) scenario assumed in Lloyd's QI system, the $n$th signal-idler pulse pair's temporal modes can be taken to be 
\begin{equation}
\frac{e^{-j(\omega_{S_o}+2\pi m/T)(t-nT_r)}}{\sqrt{T}}, 
\end{equation}
for the signal, and 
\begin{equation}
\frac{e^{-j(\omega_{I_o}-2\pi m/T)(t-nT_r)}}{\sqrt{T}}, \,
\end{equation}
for the idler, where $|t-nT_r|\le T/2$, $|m| \le (M-1)/2$, and $0 \le n \le N-1$.  Here: $\omega_{S_o}$ and $\omega_{I_o}$ are the signal and idler light's center frequencies; $M$ is an odd integer; the pulse-repetition period, $T_r$, obeys $T_r > T$; and the sign difference in the $m$-dependent part of the exponents is convenient for treating Tan~\emph{et al}.'s QI system. 

Lloyd's SP and QI analyses restrict the quantum states of the preceding temporal modes to the state space spanned by the vacuum and single-photon states.  Within this state space, his QI transmitter's $n$th signal-idler pulse pair is in the quantum state
\begin{equation}
|\psi^{(n)}\rangle_{\rm SI} = \frac{1}{\sqrt{M}}\sum_{m=-(M-1)/2}^{(M-1)/2}|{\bf 1}^{(n)}_m\rangle_S|{\bf 1}^{(n)}_m\rangle_I,
\label{LloydEntangled}
\end{equation}
where $|{\bf 1}^{(n)}_m\rangle_S$ ($|{\bf 1}^{(n)}_m\rangle_I$) denotes the $n$th signal's (idler's) state---represented as a ket vector $|\cdot\rangle_S$ ($|\cdot\rangle_I$)---containing a single photon in its $m$th mode and vacuum in its other modes.  For comparison, Lloyd's SP transmitter's $n$th signal pulse is in the quantum state
\begin{equation}
|\psi^{(n)}\rangle_S = \frac{1}{\sqrt{M}}\sum_{m=-(M-1)/2}^{(M-1)/2}|{\bf 1}^{(n)}_m\rangle_S,
\end{equation}
i.e., a superposition state containing a single photon that is equally likely to be found in any of the $M$ temporal modes.

\section{Entangled State for Tan~\emph{et al}.'s QI System \label{AppB}}
For the high time-bandwidth product ($M=TW \gg 1$) scenario assumed in Tan~\emph{et al}.'s QI system, the signal-idler pulse pair's temporal mode structure can be be taken to be 
\begin{equation}
\frac{e^{-j(\omega_{S_o} + 2\pi m/T)t}}{\sqrt{T}}, 
\label{SigFourier}
\end{equation}
for the signal, and 
\begin{equation}
\frac{e^{-j(\omega_{I_o} - 2\pi m/T)t}}{\sqrt{T}}, 
\label{IdlFourier}
\end{equation}
for the idler, where $|t|\le T/2$, and $|m| \le (M-1)/2$.  Here, as was the case for Lloyd's QI system in Appendix~\ref{AppA}, $\omega_{S_o}$ and $\omega_{I_o}$ are the signal and idler light's center frequencies, and $M$ is an odd integer.  Unlike the case for Lloyd's QI analysis, Tan~\emph{et al}.~do not restrict their system's temporal modes to the span of their vacuum and single-photon states.  In particular, the $m$th temporal modes of their signal and idler have a joint state given by
\begin{equation}
|\psi_m\rangle_{\rm SI} = \sum_{n=0}^{\infty}\sqrt{\frac{N_S^n}{(N_S+1)^{n+1}}}\,|n\rangle_{S_m}|n\rangle_{I_m},
\label{TMSV}
\end{equation}
where $|n\rangle_{S_m}$ ($|n\rangle_{I_m}$) denotes the $m$th signal mode's ($m$th idler mode's) state containing $n$ photons.  For comparison, the $m$th temporal mode of Tan~\emph{et al.}'s coherent-state transmitter is in the coherent state
\begin{equation}
|\psi_m\rangle_S =  \sum_{n=0}^{\infty}\sqrt{\frac{N_S^ne^{-N_S}}{n!}}\,|n\rangle_{S_m}.
\label{CSnumberrep}
\end{equation}

\section{Phase-Insensitive and Phase-Sensitive Cross Correlations \label{AppC}}
Because the returned light and the retained idler light are in a classical state---loss and noise having destroyed the initial entanglement of the signal and idler---we can use semiclassical photodetection theory in our analysis.  We refer to the $\langle a^*_{R_m}a_{I_m}\rangle$ cross correlation as being \emph{phase-insensitive}, because its value is invariant to applying a phase shift $\theta$ to both $a_{R_m}$ and $a_{I_m}$.  Conversely, we refer to the $\langle a_{R_m}a_{I_m}\rangle$ cross correlation as being \emph{phase-sensitive}, because its phase is shifted by $2\theta$ when a phase shift $\theta$ is applied to both $a_{R_m}$ and $a_{I_m}$.  

Photon counting can be used to measure a phase-insensitive cross correlation in a conventional interferometer.  In particular, combining the $a_{R_m}$ and $a_{I_m}$ modes on a 50--50 beam splitter we can obtain outputs
$a_{\pm_m} = (a_{R_m} \pm a_{I_m})/\sqrt{2}$.  Given hypothesis $H_j$, ideal photon counting on these outputs yield outputs whose mean values, for $j=0,1$, are
\begin{equation}
\langle |a_{\pm_m}|^2\rangle_{H_j} = \frac{\langle |a_{R_m}|^2\rangle_{H_j} \pm 2{\rm Re}[\langle a_{R_m}^*a_{I_m}\rangle_{H_j}] + \langle |a_{I_m}|^2\rangle}{2}. 
\end{equation}
For Tan~\emph{et al}.'s QI system we have $\langle a_{R_m}^*a_{I_m}\rangle_{H_j} = 0$, for $j=0,1$, so  no signature of target absence or presence is provided by the preceding interferometric measurement.  What is needed for Tan~\emph{et al}.'s QI system is a way to measure the phase-sensitive cross-correlation signature, $\langle a_{R_m}a_{I_m}\rangle_{H_j}$, for $j=0,1$.

\section{Near-Equivalence of the Quantum-Limited CS-Hom and CCN Radars \label{AppD}}
The CCN radar's source produces $M$ independent, identically-distributed, signal-idler mode pairs, $\{(\hat{a}_{S_m},\hat{a}_{I_m})\}$, that are in classically-random mixtures of coherent states whose eigenvalues, $\{(a_{S_m},a_{I_m})\}$, have the joint probability density,
\begin{equation}
p_{a_{S_m},a_{I_m}}(\alpha_S,\alpha_I) = \frac{e^{-|\alpha_S|^2/N_S}}{\pi N_S}\delta(\alpha_I-\sqrt{N_I/N_S}\,\alpha_S),
\end{equation}
where $\delta(\cdot)$ is the unit-impulse function.  In the limit $N_I \rightarrow \infty$, quantum-limited heterodyne detection of the $\{\hat{a}_{I_m}\}$ modes yields a noise-free measurement of the $\{a_{I_m}=\sqrt{N_I/N_S}\,a_{S_m}\}$.  Optimum post-heterodyne processing is then a maximal-ratio combiner~\cite{VanTrees}, which yields the \emph{conditional} ROC, given $a_{I_m} = \alpha_{I_m}$, 
\begin{eqnarray} 
\lefteqn{P_D = } \nonumber \\[.05in]
&& Q\!\left(Q^{-1}(P_F)-\sqrt{2\kappa\, \mbox{$\sum_{m={-(M-1)/2}}^{(M-1)/2}$}\frac{|\alpha_{S_m}|^2}{N_B+1}}\right),
\end{eqnarray}
where $\alpha_{S_m} = \sqrt{N_S/N_I}\,\alpha_{I_m}$.  In the limit $M \rightarrow\infty$ we have that $\sum_{m={-(M-1)/2}}^{(M-1)/2} |\alpha_{S_m}|^2 \rightarrow MN_S$, by the law of large numbers, hence the CCN radar's $N_I \gg 1$, $M\gg 1$ \emph{unconditional} ROC must satisfy
\begin{equation}
P_D \approx Q(Q^{-1}(P_F)-\sqrt{2M\kappa N_S/(N_B+1)}),
\end{equation}
which matches the CS-Hom radar's ROC when $N_B \gg 1$.  Note that the preceding near-equivalence also applies to CCN and CS-Hom radars with $N_F >1$ noise figures that satisfy $N_F\ll N_B$.


\begin{thebibliography}{99}
\bibitem{footnote1}The quantum Chernoff bound~\cite{Audenaert}, \cite{Calsamiglia}, for $N$ independent, identically-distributed observations is an exponentially-tight upper bound on the error probability, $\Pr(e)\le e^{-N\mathcal{E}}/2$, of the minimum error-probability quantum receiver, i.e., it satisfies $\lim_{N\rightarrow\infty}[\ln(\Pr(e))/N] = -\mathcal{E}$.
\bibitem{footnote4}Interestingly, the SPDC source's output state is also strongly nonclassical when $N_S \gg 1$, even though that regime is of no value to QI target detection.  That strong nonclassicality becomes observable by combining the source's signal and idler modes on a 50--50 beam splitter.  The resulting output modes, $(\hat{a}_{S_m}\pm \hat{a}_{I_m})/\sqrt{2}$, are then in single-mode squeezed states whose low-noise quadratures have variances far below the classical limit of 1/4 when $N_S \gg 1$.
\bibitem{footnote5}The high-brightness background condition for microwave operation is equivalent to $\hbar\omega \ll kT_s$, where $\omega$ is the radar frequency, $k$ is Boltzmann's constant, and $T_s$ is the sky temperature at frequency $\omega$.
\bibitem{footnote6}Thus far in this paper we have ignored---and, except here, we will continue to ignore---a target's effects on its return's polarization and frequency.  These effects, however, should be included in a more complete QI analysis.  For example, they place additional burdens---polarization tracking and Doppler compensation---on Guha and Erkmen's OPA receiver.
\bibitem{footnote7}Here, and in the next two sections, we forego consideration of QI with SFG or FF-SFG reception owing to the dim prospects for implementing those receivers with current technology.
\end{thebibliography}
\end{document}